\documentclass[twocolumn,showpacs,preprintnumbers,amsmath,amssymb]{revtex4}
\usepackage{graphicx}
\usepackage{dcolumn}
\usepackage{bm}
\usepackage{psfrag,epsfig,epsf,citesort}

\newcommand{\be}{\begin{equation}}
\newcommand{\ee}{\end{equation}}
\newcommand{\bea}{\begin{eqnarray}}
\newcommand{\eea}{\end{eqnarray}}
\newcommand{\demu}{\partial_{\mu}}

\newcommand{\de}{\partial}
\newcommand{\bear}{\begin{array}{l}}
\newcommand{\eear}{\end{array}}
\newcommand{\ds}{\displaystyle}
\newcommand{\ldl}{\Lambda \partial_{\Lambda}}
\newcommand{\inte}{\! \int \!\!}
\newcommand{\ie}{{\it i.e.}\ }
\newcommand{\cf}{{\it cf.}\ }
\newcommand{\eg}{{\it e.g.}\ }
\newcommand{\etal}{{\it et al.}}

\newcommand{\etc}{{\it etc.}\ }

\newcommand{\ug}{\!=\!}
\newcommand{\half}{\frac{1}{2}}

\def\osl{\Omega\hspace{-.25cm}/ }
\def\intoi{\int_0^\infty \!\!\! }
\def\intox{\int_0^{x} \!\!}
\def\intxi{\int_x^{\infty}\!\!\!}
\def\e{{\rm e}}
\def\lam{\lambda}
\def\Lam{\Lambda}
\def\0{\vec{0}}
\def\ker#1{\!\cdot\! #1 \!\cdot\!}
\def\hS{\hat{S}}

\def\one{\hbox{1\kern-.8mm l}}

\def\eq#1{eq.~(\ref{#1})}
\def\ceq#1{Eq.~(\ref{#1})}

\def\s#1#2#3{S^{(#1)}_{#2} (#3)}
\def\hs#1#2#3{\hS^{(#1)}_{#2} (#3)}

\def\sig#1#2#3{\Sigma^{(#1)}_{#2} (#3)}
\def\phi{\varphi}
\def\tp{\tilde{\phi}}
\def\p{\phi}
\def\tS{\tilde{S}}
\def\thS{\tilde{\hS}}
\def\tg{\tilde{\gamma}}
\def\dd{\dot{\Delta}}

\newcommand\ijmpa[3] 
		{{\it Int.\ J.\ Mod.\ Phys.\ }{\bf A #1} (#2) #3}

\newcommand\jhep[3]  
		{{\it J. High Energy Phys.\ }{\bf #1} (#2) #3}

\newcommand\npb[3]   
		{{\it Nucl.\ Phys.\ }{\bf B #1} (#2) #3}

\newcommand\pla[3]   
		{{\it Phys.\ Lett.\ }{\bf A #1} (#2) #3}
\newcommand\plb[3]   
		{{\it Phys.\ Lett.\ }{\bf B #1} (#2) #3}

\newcommand\preva[3]  
		{{\it Phys.\ Rev.\ }{\bf A #1} (#2) #3}

\newcommand\prevd[3]   
		{{\it Phys.\ Rev.\ }{\bf D #1} (#2) #3}

\newcommand\revmp[3]   
	{{\it Rev.\ Mod.\ Phys.\ }{\bf #1} (#2) #3}

\newcommand\tmp[3]   
		{{\it Theor.\ Math.\ Phys.\ }{\bf #1} (#2) #3}

\newcommand\hepph[1] {{\tt hep-ph/#1}}
\newcommand\hepth[1] {{\tt hep-th/#1}}

\begin{document}

\preprint{CERN-CH/2003-264}
\preprint{SHEP 03-17\phantom{CH/264}}
\title{Exact scheme independence at two loops}

\author{Stefano Arnone}\email{stefano.arnone@roma1.infn.it}
\affiliation{Department of Physics and Astronomy,
University of Southampton\\
Highfield, Southampton SO17 1BJ, U.K.\\
and Dipartimento di Fisica\\
Universit\`a di Roma ``La Sapienza''\\
P.le Aldo Moro, 2 - 00185 Roma - Italy}
\author{Antonio Gatti}\email{A.Gatti@damtp.cam.ac.uk}
\affiliation{DAMTP, Centre for Mathematical Sciences,\\
Wilberforce Road, Cambridge CB3 0WA, U.K.}

\author{Tim R. Morris}\email{T.R.Morris@soton.ac.uk}
\altaffiliation[Currently at ]{CERN, Theory Division, CH-1211 Gen\`eve 23,
Switzerland}
\author{Oliver J. Rosten}\email{O.J.Rosten@soton.ac.uk}
\affiliation{Department of Physics and Astronomy,
University of Southampton\\
Highfield, Southampton SO17 1BJ, U.K.}

\date{\today}

\begin{abstract}
We further develop an algorithmic and diagrammatic
computational framework for very general exact renormalization
groups, where the embedded regularisation scheme, parametrised by
a general cutoff function and infinitely many higher point
vertices, is left unspecified. Calculations proceed iteratively,
by integrating by parts with respect to the effective cutoff, thus
introducing effective propagators, and differentials of vertices
that can be expanded using the flow equations; many cancellations
occur on using the fact that the effective propagator is the
inverse of the classical Wilsonian two-point vertex.
We demonstrate the power of these methods by computing the beta
function up to two loops in massless four dimensional scalar field
theory, obtaining the expected universal coefficients, independent
of the details of the regularisation scheme. 
\end{abstract}
\maketitle

\section{Introduction and conclusions}

The deeper understanding of renormalization, due to Wilson,
follows most directly in the continuum from the exact
renormalization group (ERG) flow equations \cite{Wil}. The fact
that solutions of these equations, for the Wilsonian effective
action $S$, can be found directly in terms of renormalised
quantities, that all physics (\eg Green functions) can be
extracted from $S$, and that renormalizability is trivially
preserved in almost any approximation \cite{morig,rev}, turns
these ideas into a powerful framework for considering both
perturbative and non-perturbative approximations (see for example
refs.
\cite{Bo1,Bo,Kopietz:2000bh,LM,Mo,Mo1,Papenbrock:kf,Pernici:1998tp,Po,Tighe,We,Wil,Zappala:2002nx,ant,ball,beta2g,eqs,general,has,morig,oneloop,rev,rg2002,sumi,truncam,wein}).

In the past a number of different versions and ways of deriving
the ERG have been proposed
\cite{Wil,morig,rev,has,general,wein,Po,We,Bo,Mo1}, which however
have been shown to be equivalent under changes of variables
\cite{morig,rev,truncam,sumi,LM}. Recently, far more general
versions of the ERG have been considered \cite{LM}. All the ERGs,
including these generalised ones, can be seen to be parametrised
by a functional $\Psi$ \cite{Mo1,LM}, that induces a
reparametrisation (field redefinition) along the flow, and acts as
a connection between the theory space of actions at different
effective cutoff scales $\Lambda$. As a result, local to some
generic point $\Lambda$ on the flow, all these ERGs may be shown
to be just reparametrisations of each other. When this
reparametrisation can be extended globally, the result is an
immediate proof of scheme independence for physical observables.
Indeed computations of physical quantities then differ only
through some field reparametrisation.

One practical example is an explicit field redefinition that
interpolates between results computed using different choices of
cutoff function \cite{LM}. Even more dramatic than this however,
is the use of this freedom to adapt the ERG to certain forms of
approximation or special physical problems \cite{LM}. In
particular, recently there has been substantial progress in
adapting these ideas to gauge theory. It turns out that not only
can one introduce an effective cutoff $\Lambda$ in a way that does
not break the gauge invariance \cite{gareg} but careful choices of
$\Psi$ allow the gauge invariance to be preserved manifestly, in
fact not even gauge fixed, along the flow and in the solutions $S$
\cite{Mo1,Mo,ant}.

Ref.\  \cite{LM} did not answer the question of precisely when the
automatic local equivalence of two ERGs can be extended globally.
In ref.\  \cite{oneloop}, we showed that for four dimensional one-component 
scalar field theory, the universal one-loop $\beta$
function is obtained for a general form of $\Psi$,  involving a
very general `seed' action $\hS$. For simplicity we chose to keep
$\phi\leftrightarrow-\phi$ invariance ($\phi$ being the scalar
field) and the $\hS$ two-point vertex was specified to be equal to
that of the classical effective action, which thus determines
them, up to a choice of cutoff function $c$. The only further
requirements we imposed were that {\it the vertices of $\hS$ be
infinitely differentiable and lead to convergent momentum
integrals}.  These easily met requirements are needed in any case
for the ERG equation to make sense at the quantum level
\cite{oneloop}.

In this paper we will show that the universal two-loop $\beta$
function also comes out correct for this very general form of
$\Psi$. It is natural to conjecture then, that to all orders in
perturbation theory, the only constraints required on $\hS$ to get
universal answers for physical quantities are the ones in italics
above. Furthermore, it is surely possible to show in perturbation
theory, that all such ERGs are reparametrisations of each other,
and indeed to construct the map perturbatively.

We will see that the iterative diagrammatic computational
framework introduced in ref.\  \cite{oneloop}, and further refined
in the gauge theory context in ref.\  \cite{ant}, extends
straightforwardly to higher loops. The method works by turning the
large redundancy in $\Psi$ (here encapsulated in $\hS$) to our
advantage. Since we are not allowed to inquire into the form of
its vertices, the calculational steps are severely limited. The
inherent generality thus acts as a roadmap to the most efficient
computation. Indeed, since the form of the vertices is never
specified, the majority of the calculation is best performed by
manipulating the diagrams themselves.

However, it should be emphasised that, although the calculational
procedure is the most efficient, the actual computation of the
two-loop $\beta$-function, presented here, could be considerably
shorter. Indeed, if our purpose were just to compute the two-loop
$\beta$-function then, having used the redundancy of $\Psi$ to uncover
the best calculational procedure, we could use the simplest form of
$\hat{S}$ sufficient to yield a valid ERG. This corresponds to
discarding all seed action vertices, other than the classical
two-point vertex, reducing the ERG equation to Polchinski's
version. We would then be left with a powerful algorithmic and
diagrammatic procedure, which does not require the cutoff function $c$
to be specified.

Our aim is rather to explicitly demonstrate scheme independence and so
we do not restrict $\hat{S}$ in this way. Indeed, we even allow
$\hat{S}$ to have its own completely unspecified (loop) expansion in
$\hbar$. Furthermore, by retaining a generic seed action, we gain
valuable experience for the gauge theory case, where $\hat{S}$
necessarily contains many interaction vertices~\cite{ant}.

An important prerequisite for this method to work, is that the
effective action is written in self-similar form \cite{shirkov},
in which no other explicit scale is introduced except $\Lambda$.
(Note that although we formulate the method for a massless theory,
nothing precludes the method from being applied to massive
theories. The required self similar flow is achieved in this case
simply by introducing the appropriate RG concept of a running
mass, \ie a function of $\Lambda$ with its own $\beta$ function,
see for example \cite{eqs,rev}.)

We also constrain the form of the flow equation so that in this
context, the kernel appearing in the ERG equation, is the
differential of the effective propagator, the latter being the
inverse of the classical effective action two-point vertex. (In
gauge theory this condition holds only up to gauge transformations
\cite{ant,rg2002}.)

The computation then proceeds as follows. We introduce effective
propagators in diagrams containing no explicit seed action
vertices, by integrating by parts with respect to $\Lambda$. This
results in total $\Lambda$-derivative contributions, and terms
where the $\Lambda$-derivative acts on effective action vertices.
We use the flow equation to process these latter further. Then,
using the fact that the effective propagator is the inverse of the
classical two-point vertex, many cancellations typically occur.
The above procedure is repeated until there is no explicit
dependence left upon the (generic) seed action $\hS$. The whole of
this iterative procedure may be performed entirely
diagrammatically.

Universal terms  arise from the total $\Lambda$-derivative
contributions. Whenever the total $\Lambda$-derivative acts on
convergent momentum integrals, the result follows trivially by
dimensions; in particular, we frequently exploit the fact that
dimensionless cases simply vanish. Although ultraviolet finiteness
is built in to the ERG, non-vanishing universal terms arise from
dimensionless momentum integrals that are infrared convergent only
{\sl after} the $\Lambda$-derivative is taken. At the two-loop
level, we generate for the first time, contributions that are
infrared divergent even after differentiation with respect to
$\Lambda$. Of course at finite $\Lambda$, everything is infrared
finite in the ERG, again by construction. Thus the game here is to
rearrange so these infrared divergences cancel amongst each other;
intelligent combinations of these terms then are convergent and
vanish, or at worst result in calculable universal contributions.

In this way, the standard two-loop $\beta$ function coefficient,
$-17/3(4\pi)^4$, is derived, however without specifying at any
stage the form of the cutoff function or the form of any of the
higher point vertices.

We note that this two-loop $\beta$ function coefficient has
already been derived within the ERG, with various cutoff
functions, and in various ways, corresponding to differing
motivations
\cite{Papenbrock:kf,Pernici:1998tp,Zappala:2002nx,Tighe,Bo,Kopietz:2000bh}.
Although ref.\  \cite{Bo} considered a general cutoff function, this
work is the first to treat the more general case of an arbitrary
seed action $\hS$ (corresponding to a continuum version of an
arbitrary ``blocking scheme'' \cite{Wil}), and to reduce the
computation in this general framework to a largely algorithmic and
diagrammatic approach. We expect these insights to be especially
useful for higher loop computations in the gauge invariant ERGs
\cite{beta2g}.

\section{\label{sec:ag} A generalised exact renormalization group equation}

We will consider a massless one-component scalar field theory in
four Euclidean dimensions, 
and sketch the derivation of a generalised ERG equation, starting
from Polchinski's equation~\cite{Po}. For details we refer the reader
to~\cite{oneloop}.

The Wilsonian renormalization group (RG) is defined in terms of
some effective ultraviolet cutoff $\Lam$~\cite{Wil}. Polchinski's
version of Wilson's ERG equation~\cite{Wil} implements this
transparently through a cutoff function $c(p^2/\Lambda^2)$, which
modifies propagators $1/p^2$ to $\Delta = c/p^2$. We take $c$ to
be {\it smooth}, \ie infinitely differentiable, and always
positive. It satisfies $c(0)=1$ so that low energies are
unaltered, and tends to zero as $p^2/\Lambda^2\to\infty$
sufficiently fast that all Feynman diagrams are ultra-violet
regulated.

The partition function is given as the functional integral of the
measure, ${\rm e}^{-S}$, where $S$ is the Wilsonian effective
action
\be
\label{wilsact} 
S[\varphi;\Lam] = \half \inte {d^4 p \over
(2\pi)^4} \, p^2 c^{-1}_p \, \varphi^2 + S^{int}[\varphi;\Lam],
\ee 
$c^{-1}_p \equiv c^{-1}(p^2/ \Lam^2)$. The first term in the
above, namely the regularised kinetic term, will be referred to as
the seed action and denoted by $\hS$.

Demanding that physics be invariant under the renormalization
group transformation $\Lam\mapsto\Lam-\delta\Lam$, results in a
functional differential equation for the effective interaction
\cite{Po}, which can be recast in terms of the total effective
action, $S=\hS+S^{int}$ [\cf \eq{wilsact}], $\Sigma \doteq
S-2\hS$, and the differentiated effective propagator  $\dd \doteq
-\ldl\Delta ={2\over\Lam^2} \, c'(p^2/\Lam^2)$, as
 \be
 \label{pofleq} \dot{S} \equiv - \ldl S =
{1\over 2}
 {\delta S\over\delta\varphi} \ker{\dd}
{\delta \Sigma \over\delta\varphi} - {1\over 2}{\delta
\over\delta\varphi} \ker{\dd} {\delta \Sigma \over\delta\varphi}
 \ee
(up to a vacuum energy term discarded in~\cite{Po}. We
will also refer to $\dd$ as the kernel.)

In the above, prime denotes differentiation with respect to the
function's argument (here $p^2/\Lambda^2$) and the following shorthand has
been introduced: for any two functions $f(x)$ and $g(y)$ and a
momentum space kernel $W(p^2/\Lambda^2)$,
with $\Lam$ being the effective cutoff,
\be
f \ker{W} g =
\int\!\!\!\!\int\!\!d^4\!x\,d^4\!y\
f(x)\, W_{x y}\,g(y),
\ee
where $W_{x y} = \inte {d^4 p \over (2\pi)^4} \, W(p^2/\Lambda^2) {\rm
e}^{i p \cdot (x-y)}$.
($\hS$ may therefore be written as $\half\,\demu \varphi \ker{c^{-1}}
\demu\varphi$.)

\begin{figure}
\includegraphics{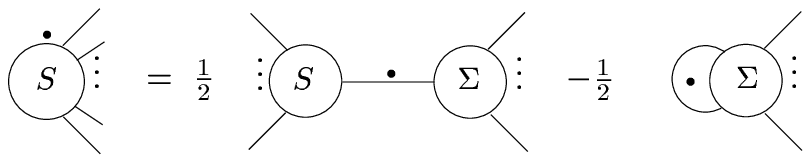}
\caption{\label{fig:rgeq}Graphical representation
of the ERG equation. $S, \Sigma$ $n$-point vertices are
represented by circles labelled respectively by $S, \Sigma$ with
$n$ legs attached, while bulleted lines stand for $\dd$.}
\end{figure}

\ceq{pofleq}, which is given diagrammatically as in
fig.~\ref{fig:rgeq}, can be turned into a flow equation for
the measure,  
\be \label{measurefl} \ldl \e^{-S} = -{1\over
2}{\delta\over\delta\varphi}\cdot \dd \cdot \left( {\delta\Sigma
\over\delta\varphi}\,\e^{-S}\right), \ee which makes the
invariance of the partition function under the renormalization
group transformation manifest by showing that the measure flows
into a total functional derivative.

Eqs.~(\ref{pofleq}), (\ref{measurefl}) may also be reinterpreted
in a slightly different way: introducing $\Psi = -{1\over 2 \Lam}
\, \dd \, {\delta \Sigma \over\delta\varphi}$~\cite{LM}, we can
regard the change in the partition function as corresponding to
the field redefinition $\varphi \rightarrow \varphi + \delta \Lam
\, \Psi$, from which we can infer that integrating out degrees of
freedom is just equivalent to a reparametrisation of the partition
function~\cite{Mo,LM}. This somewhat counterintuitive result is
allowed in the continuum because of the infinite number of degrees
of freedom per unit volume.

Different forms of ERG equations are obtained by choosing
different $\Psi$, and there is a great deal of freedom in such a
choice of ``blocking scheme''~\cite{LM}. Nonetheless, physical
quantities should not depend on the particular scheme, in other
words they should turn out to be universal.  Therefore we can
choose a $\Psi$ 
which is best suited for our purposes (for example one that
generates a manifestly gauge invariant ERG~\cite{Mo,ant}) and
still get the right answer when computing physical quantities.

Moreover, if physical results are to come out the same,
irrespective of the choice of $\Psi$, we might well decide not to
pick a specific form for it, but rather leave it general (except
for some very general requirements discussed later) and implement
a method for calculating universal quantities which does not
depend on the details put in by hand. The main advantage
of such a procedure is that we must be able to see that all the dependence of
our results on unspecified quantities clearly cancels out.

In this paper we want to go along these lines and generalise our
previous results~\cite{oneloop} by calculating the two-loop beta
function in a scalar field theory with a generalised $\Psi$. With
so much freedom we have to restrict it to be able to be concrete;
we choose to consider a general seed action $\hS$ of the form
outlined below.

Firstly, we want the tree-level two-point vertex in $\hS$ to be the same as
the required two-point vertex for the classical effective action.
Since we are dealing with a massless theory, that means that both
two-point vertices should be set to be equal to the regularised
kinetic term in \eq{wilsact}\footnote{Thus no further bilinear
term must be hiding in $S^{int}$, at the classical level.}. We will
see in the next section that this constraint can be consistently
imposed on solutions of the flow equation (\ref{pofleq}).

(In general we impose that the two-point vertices of $\hS$ and the
effective action must coincide at the classical level. The form of the flow
equations then imply that $\dd$ really is the differentiated
propagator in this context, {\it i.e.\ \emph{the}} differential of
$\Delta$, where $\Delta$ is the inverse of the classical effective
action two-point vertex. Here we already solved this constraint in
writing \eq{pofleq}. However, for a massive theory, any classical
mass would have to appear already in $\Delta$, thus slightly
generalising \eq{pofleq}. Furthermore, in manifestly gauge
invariant ERGs~\cite{Mo,ant}, the inverse of the classical
two-point vertex does not exist for the gauge fields; instead,
$\Delta$ is the inverse, only up to a gauge transformation. It is
not necessary that the two-point vertices of $\hS$ and the
classical effective action coincide. We choose to require this
purely for the significant technical advantages it brings to our
method, as described later.)

Secondly, we choose to leave the $\phi\leftrightarrow-\phi$
symmetry alone, which implies that $\hS$ must be even under this
symmetry. We are left with a generalised exact renormalization
group parametrised by the infinite set of seed action $2n$-point
vertices, $n \geq 2$. We will leave each of these vertices as
completely unspecified functions of their momenta except for the
very general requirements that {\it the vertices be infinitely
differentiable and lead to convergent momentum integrals}. The
first condition ensures that no spurious infrared singularities
are introduced and that all effective vertices can be Taylor
expanded in their momenta to any order \cite{Mo1,rev}. The second
condition is necessary for the flow equation to make sense at the
quantum level and also ensures that the flow actually corresponds
to integrating out modes \cite{Mo,LM}.

We are therefore incorporating in the momentum dependence of {\sl
each} of the seed action $2n$-point vertices an infinite number of
parameters. Of course these infinite number of vertices, each with
an infinite number of parameters, then appear in the effective
action $S$ as a consequence of the flow equation. Remarkably, we
can still compute the two-loop $\beta$ function.

\section{Preliminaries}
\subsection{Self-similar flow}

We denote the vertices of the effective action as
\be
\s{2n}{}{\vec{p};\Lam}\equiv\s{2n}{}{p_1,p_2,\cdots,p_{2n};\Lam},
\ee where we have factored out the momentum conserving $\delta$
function:
\be
(2\pi)^{4} 
\delta\!\left(\!\sum^{2n}_{i=1}p_i\!\right)\,
\s{2n}{}{\vec{p};\Lam}
 \doteq (2\pi)^{8n-4} {\delta^{2n}
S\over\delta\p(p_1)
\cdots\delta\p(p_{2n})} \ee (and
similarly for $\hS$).

In a standard perturbative treatment, we would define the theory,
by stating that at the classical level the Lagrangian density
takes the form
\be
{1\over2}(\partial_\mu\varphi)^2 + {\lam \over 4!} \varphi^4.
 \ee
After regularising the theory, the coupling is replaced by a bare
coupling, and a bare mass is introduced, and these are solved for,
order by order in $\lam$, so that at physical scales all
quantities of physical interest ({\it e.g}.\ Green functions) are
finite and correspond to a massless theory. An essential step in
this treatment is to define what we really mean by $\lam$ through
some renormalization condition.

If we regularise the theory, by using the cutoff function $c$ at
some ultra-violet cutoff scale $\Lam=\Lam_0$, then the Wilsonian
effective action at physical scales $S[\varphi;\Lam]$ will simply
result from integrating \eq{pofleq}, with the initial condition
that $S[\varphi;\Lam_0]$ is the bare action. This can be done
order by order in $\lam$. In the continuum limit
$\Lam_0\to\infty$, we obtain the Wilsonian effective action
expressed in renormalized terms.

We will not approach the calculation this way. Instead, firstly we
choose renormalization conditions that only involve the scale
$\Lam$, for example in this case we use
 \be 
\s{2}{}{p;\Lam} \equiv \s{2}{}{p,-p;\Lam} =
\sigma(\lam)\Lam^2 + p^2
  +O(p^4/\Lam^2),
\label{or2} 
\ee
\be
\s{4}{}{\vec{0};\Lam} = \lam,\label{or4} 
\ee
where $\sigma=\sigma_1\lam+\sigma_2\lam^2+\cdots$ is a function
of $\lam$ which we determine self-consistently. This sets the
physical mass to zero implicitly by ensuring that the only scale
that appears is $\Lam$, which itself tends to zero as all momenta
are integrated out \footnote{If $\sigma$ has an $O(\lam^0)$
coefficient, the only other possibility arises, namely that the
physical mass is infinite. This will be dealt with in the next
gauge theory paper.}. In order to obtain unit $p^2$ coefficient in
\eq{or2}, of course we also have to reparametrise the field by the
wavefunction renormalization factor. In the limit that
$\Lam_0\to\infty$, apart from the $\Lam$ dependence expected by
na\"\i ve ({\it a.k.a.\ \emph{engineering}}) dimensions, the only
dependence on $\Lam$ then appears through $\lam(\Lam)$.
Equivalently, if we rescale the field and positions/momenta by the
appropriate (engineering) powers of $\Lam$ to make everything
dimensionless, then the effective action takes the self-similar
\cite{shirkov} form $S[\varphi;\lam]$,  in which {\sl the only
dependence on $\Lam$ is through the coupling} $\lam(\Lam)$. The
existence of such a solution is equivalent to a statement of
renormalizability, and the solution corresponds to the
renormalized trajectory \footnote{as usual ignoring in perturbation
theory, the triviality problems of scalar field theory}
\cite{rev}.

Secondly, we considerably simplify the analysis, both conceptually
and in terms of procedure, by solving the ERG directly in terms of
the self-similar solution $S[\varphi;\lam]$, 
order by order in $\lam$
\cite{morig,rev,Mo1,Mo,ant,Tighe,oneloop,eqs}. In this way, we
never introduce an overall cutoff  ({\it e.g.\ $\Lam_0$}) or the
notion of a bare action at all. Instead, we obtain the continuum
physics directly.

\subsection{Perturbative expansion}

Rather than directly use the form of the flow equation in \eq{pofleq},
we first make explicit the wavefunction renormalization contribution.
Moreover, we rescale $\phi$ so as to put the coupling constant in
front of the action. This will ensure the expansion in the coupling
constant coincides with the one in $\hbar$, the actual expansion
parameter being in fact $\lambda \hbar$. Our flow equation then
reads~\cite{oneloop}
\be \label{tildefleq} 
\bear
{\ds 
- \ldl \left({1\over
\lam} \tS\right) + {\tg\over2\lam} \int_p
 \tp(p){\delta \tS\over \delta \tp(p)} 
}\\[0.5cm]
{\ds
=\frac{1}{2\lambda}\frac{\delta (\tS-2\thS)}{\delta\tp} \ker{\dd}
\frac{\delta \tS}{\delta \tp}-\frac{1}{2}\frac{\delta}{\delta
\tp}\ker{\dd}\frac{\delta(\tS-2\thS)}{\delta\tp}, 
}
\eear
\ee 
where
$\int_p \equiv \inte {d^4 p\over (2\pi)^4}$ and a tilde has been
put over rescaled quantities---so for example $\tp \equiv
\frac{1}{\sqrt{\lam}} \phi$ and $\tS \equiv \frac{1}{\lam} S$---as
well as over the anomalous dimension $\gamma$ ($\gamma = \ldl \log
Z$, $Z$ being the wavefunction renormalization) to signify it has
absorbed the change to $\left. \ldl \right|_{\tp}$.

(Note that \eq{tildefleq} is actually not the result of
rescaling the wavefunction renormalization out of \eq{pofleq}.  We
must in addition change the cutoff function $c\mapsto cZ$ in the flow
equation~\cite{oneloop,Mo1,LM}.  However, the important point is that
it clearly still satisfies the requirements of an ERG equation: namely
that the partition function is invariant under the flow and that the
flow corresponds to integrating out degrees of
freedom~\cite{ant,LM}. Hence it is a valid and even more appropriate
starting point when the wavefunction renormalization is to be taken
into account. This is another example of the immense freedom in the
choice of the ERG equation.)

In order to simplify the notation, the tildes will be removed from
now on. In these new, rescaled variables, the renormalization
conditions are set to
 \be
\s{2}{}{p;\Lam} \equiv \s{2}{}{p,-p;\Lam} =
\sigma(\lam)\Lam^2 + p^2
  +O(p^4/\Lam^2),
\label{r2} 
\ee
\be
\s{4}{}{\vec{0};\Lam} = 1.\label{r4} 
 \ee
Both conditions are already saturated at tree level. (To see this
it is sufficient to note that, since the theory is massless, the
only scale involved is $\Lam$. Since the tree-level four-point vertex  
is dimensionless, it must be a constant at null momenta. Thus 
$\s{4}{0}{\vec{0};
\Lam} = \s{4}{0}{\vec{0}; \Lam_0} = 1$, where the lower index $0$ signifies
the coupling is intended at the tree level. Similar 
arguments apply to the tree-level two-point function.)

Expanding the action, the seed action, the beta 
function $\beta(\Lam) = \ldl \lam$
and the anomalous dimension in powers of the coupling constant,
and bearing in mind that the (inverse of the) coupling constant
now appears in front of $S$: 
\bea \nonumber S[\p;\lam]&=&S_0+\lambda S_1+\lambda^2
S_2+\cdots,\\ \nonumber
\nonumber \hS[\p;\lam]&=&\hS_0+\lambda \hS_1+\lambda^2
\hS_2+\cdots,\\ \nonumber
\beta(\Lam)&=&\beta_1\lambda^2+\beta_2\lambda^3+\cdots,\\
\nonumber \gamma(\Lam) &=&\gamma_1\lam+\gamma_2\lam^2+\cdots 
\eea
yields the loopwise expansion of the flow equation 
\begin{widetext}
\bea \dot{S}_0
&\!=\!&\frac{1}{2}\frac{\delta S_0}{\delta\varphi} \ker{\dd}
\frac{\delta \Sigma_0 }{\delta\varphi},\label{scalartree}\\
 \dot{S}_1+\beta_1 S_0+{\gamma_1\over2}\ \phi\!\cdot\!
{\delta S_0\over\delta\phi} &\!=\!&
\frac{\delta S_1}{\delta\varphi}\ker{\dd}\frac{\delta \Pi_0 }
{\delta\varphi}
- \frac{\delta S_0}{\delta\varphi}\ker{\dd}\frac{\delta \hS_1 }
{\delta\varphi} 
-\frac{1}{2} \frac{\delta}{\delta\varphi}\ker{\dd}
\frac{\delta \Sigma_0 }{\delta\varphi},\label{scalar1loop}\\
\dot{S}_2+\beta_2 S_0 + {\gamma_1\over2}\ \phi\!\cdot\!
{\delta S_1\over\delta\phi} + {\gamma_2\over2}\ \phi\!\cdot\!
{\delta S_0\over\delta\phi} &\!=\!&
\frac{\delta
S_2}{\delta\varphi}\ker{\dd}\frac{\delta \Pi_0 }{\delta\varphi}
- \frac{\delta S_0}{\delta\varphi}\ker{\dd}\frac{\delta \hS_2 }
{\delta\varphi}
+ \frac{1}{2} \frac{\delta S_1}{\delta\varphi}\ker{\dd}\frac{\delta
\Sigma_1}{\delta\varphi} -\frac{1}{2} \frac{\delta}{\delta\varphi}\ker{\dd}
\frac{\delta \Sigma_1}{\delta\varphi}\label{scalar2loop}\\ \nonumber
\eea 
\end{widetext}
\etc where again dots above quantities signify $-\ldl$,
$\Sigma_n = S_n-2\hS_n$ and $\Pi_n = S_n-\hS_n$.

$\beta$ and $\gamma$, at one- and two-loop order, may be
extracted directly from
eqs.~(\ref{scalar1loop}),(\ref{scalar2loop}), as specialised to the two-
and four-point effective couplings, 
once the renormalization conditions have been taken into account.

The procedure is very straightforward: as the renormalization
conditions are already saturated at tree level, there must be no
quantum corrections to the two-point effective coupling at order
$p^2$, nor to the four-point at null momenta, \ie \be
\label{rencon} \left. \s{2}{n}{p;\Lam} \right|_{p^2} =
\s{4}{n}{\vec{0};\Lam} = 0 \qquad \forall n \geq 1, \ee where the
notation $\left.f \right|_{p^2}$ signifies that the coefficient of
$p^2$ in the series expansion of $f$ must be taken. Hence the ERG
equations for these special parts of the quantum corrections
greatly simplify, reducing to algebraic equations which can be
solved for $\beta, \gamma$ at any order of perturbation theory.

As explained in the previous section, our $\hS$ is not completely
arbitrary. Apart from the very general requirements on the
differentiability and integrability of its vertices mentioned
earlier, for convenience we restrict $\hS$ to have only even-point
vertices, and constrain its tree-level two-point vertex so that it is equal
to that of the classical effective action:
 \be
 \label{shat20} \hS_0^{(2)}(p) = \s{2}{0}{p}.
 \ee
Although this constraint is not necessary, it greatly simplifies
the flow equations for higher point vertices, as it implies, for
example, that $\Pi_0^{(2)} \equiv 0$. Recall that we also wish to
set
 \be
 \label{s20} \s{2}{0}{p} = p^2 c^{-1}_p.
 \ee
However, $\s{2}{0}{p}$ is already determined, up to an integration
constant, by the flow equation. Using \eq{shat20} and the
two-point part of \eq{scalartree}, and rearranging, we have
 \be
 \label{s20consistent}
 \ldl \left(\s{2}{0}{p}
 \right)^{-1} = \ldl \Delta_p,
 \ee
which shows that \eq{s20} is consistent with the flow equation.
Indeed,
 \be
 \label{intwin}
 \s{2}{0}{p}\, \Delta_p = 1,
 \ee
a central relation in the calculation that follows.

\begin{figure}[!h]
 \includegraphics{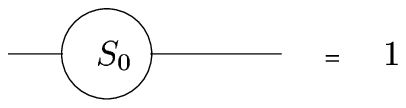}
\caption{Diagrammatic representation of
 \eq{intwin}.\label{fig:intwin}}
\end{figure}
Before going to the details of the two-loop calculation, to which
the next section will be devoted, we will describe what our method
consists of by rederiving the one-loop contribution to
$\beta$~\cite{oneloop}. Being a much simpler calculation, it
constitutes the perfect ground for illustrating our strategy and,
moreover, it will help the reader to get familiar with
the diagrammatics.

In order to ease notation in the figures, 
in what follows the vertices of the effective action will be labelled by
their loop order only, and those of $\hS$ by their loop order with a ``hat''
on top. 

\subsection{One-loop beta function} 

\begin{figure}[!h]
\includegraphics{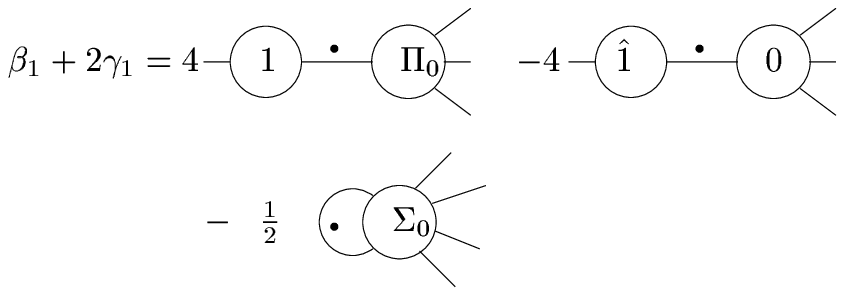}
\caption{Graphical representation of \eq{beta1}.
All the external legs have null momentum.\label{fig:beta1}}
\end{figure}

\begin{figure}[!h]
\includegraphics{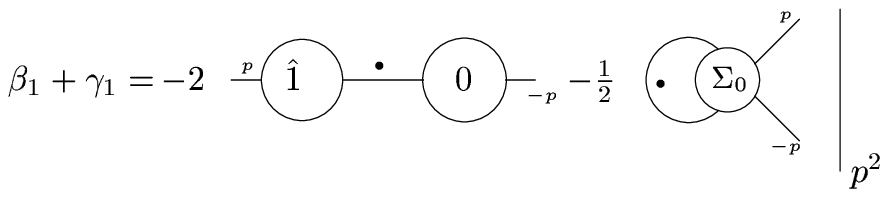}
\caption{Graphical
representation of \eq{gamma1}.\label{fig:gamma1}}
\end{figure}

Specialising \eq{scalar1loop} to the two- and four-point effective
couplings 
and imposing the renormalization conditions, \eq{rencon}, yields
\bea 
\beta_1+2\gamma_1 &\!=\!& 4 \s{2}{1}{0} \, \dd_0 \, \Pi_0^{(4)} (\vec{0})
- 4 \s{4}{0}{\vec{0}}\, \dd_0 \, \hS_1^{(2)} (0) \nonumber\\ 
&-& \frac{1}{2}\int_q \, \dd_q \, \Sigma_0^{(6)}(\vec{0},q,-q), \label{beta1}
\eea
\bea
\beta_1+\gamma_1 &\!=\!& \left. -2 \s{2}{0}{p}\, \dd_p \, \hS_1^{(2)} (p)
\right|_{p^2} \nonumber\\
&-& \frac{1}{2}\int_q \, \dd_q \, \left. \Sigma_0^{(4)}
(p,-p,q,-q)\right|_{p^2}, \label{gamma1} 
\eea
where $\dd_0 = \frac{2}{\Lam^2} c'(0)$.
Eqs.~(\ref{beta1}), (\ref{gamma1}) are represented diagrammatically as in
figs.~\ref{fig:beta1}, \ref{fig:gamma1}.


We start processing \eq{gamma1} by integrating by parts the diagram
containing the four-point $S_0$ vertex. 
\be \label{1step}
\bear
{\ds \beta_1+\gamma_1 = - \frac{1}{2}\int_q \left[ \Delta_q \,
\s{4}{0}{p,-p,q,-q} \right]_{p^2}^{\scriptscriptstyle \bullet} + 
\frac{1}{2}\int_q \, \Delta_q }\\
{\ds \left. \times \dot{S}_0^{(4)} (p,-p,q,-q) \right|_{p^2}
+ \int_q \, \dd_q \, \left. \hS_0^{(4)}
(p,-p,q,-q)\right|_{p^2} }\\
{\ds\left. -2 \hS_1^{(2)}(p) \, \dd_p \, \s{2}{0}{p}\right|_{p^2}. }
\eear 
\ee   
\begin{figure}[!h]
\includegraphics{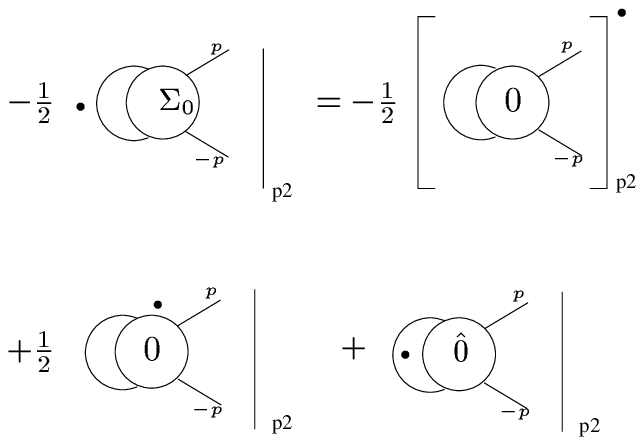}
\caption{Integrating by parts the four-point vertex in
\eq{gamma1} [\cf \eq{1step}].\label{fig:intparts}}
\end{figure}

Now, the first term in the above represents a fully convergent integral
[\cf \eq{1step}],
therefore the order of the derivative and integral signs can be
exchanged. Moreover, as the integrand is dimensionless, there can be no
dependence upon $\Lam$ after the momentum integral is carried out. Hence
the result vanishes identically. The second term
can be processed further by making use of the tree-level flow equation
for $\s{4}{0}{p,-p,q,-q}$:
\be
\bear 
{\ds 
\frac{1}{2} \left. \Delta_q \, \dot{S}_0^{(4)} (p,-p,q,-q)
\right|_{p^2} = \frac{1}{2}
\Delta_q \Big[ \hS_0^{(4)} (p,-p,q,-q) } \nonumber \\
{\ds \times \Big( -2 \s{2}{0}{p} \, \dd_p
-2 \s{2}{0}{q} \, \dd_q \Big) \Big]_{p^2} = - \dd_q  }\nonumber \\
\eear
\ee
\be
\bear \label{2step}
{\ds \times \left. \hS_0^{(4)} (p,-p,q,-q)\right|_{p^2} 
- \Delta_q \,
\s{2}{0}{p} \, \dd_p \left. \hS_0^{(4)}(p,-p,q,-q) \right|_{p^2}. }
\eear
\ee
\begin{widetext}
\phantom{ABCDE}
\vspace{-1em}
\begin{figure*}[!h]
\includegraphics{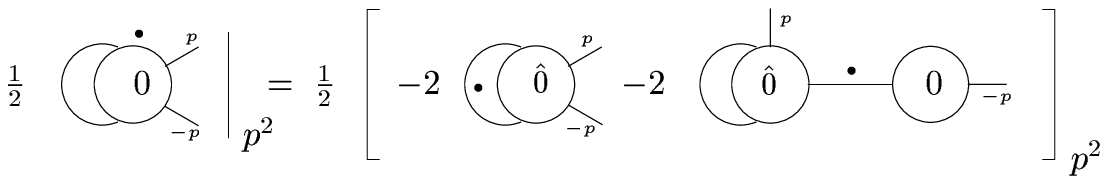}
\caption{Using the flow equation for
$S_0^{(4)}$.\label{fig:4pteq}}
\end{figure*}
\begin{figure*}[!h]
\includegraphics{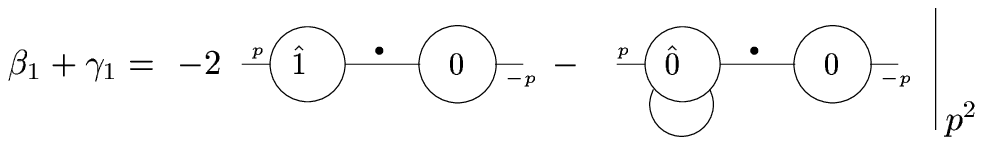}
\caption{Eq.~(\ref{gamma1}) in its final form.\label{fig:b1g1mod}}
\end{figure*}
\end{widetext}

Using the relation $\s{2}{0}{q} \Delta_q = 1$ (\cf fig.~\ref{fig:intwin}), 
one of the two terms on the
r.h.s. has been simplified. We can see that it just cancels the four-point 
$\hS$ contribution in
\eq{1step}. The remaining one will cancel when eqs.~(\ref{beta1}),
(\ref{gamma1}) are solved for $\beta_1$. \ceq{2step} is shown in 
fig.~\ref{fig:4pteq}, 
while \eq{gamma1} in its final form is displayed as in 
fig.~\ref{fig:b1g1mod}.  

We can diagrammatically process \eq{beta1} in pretty much the same
way: integrating by parts and making use of the flow equation for
the six-point effective coupling, we get the diagrams shown in
fig.~\ref{fig:6pteq}. 
\begin{widetext}
\phantom{ABCDE}
\begin{figure*}[!h]
\includegraphics{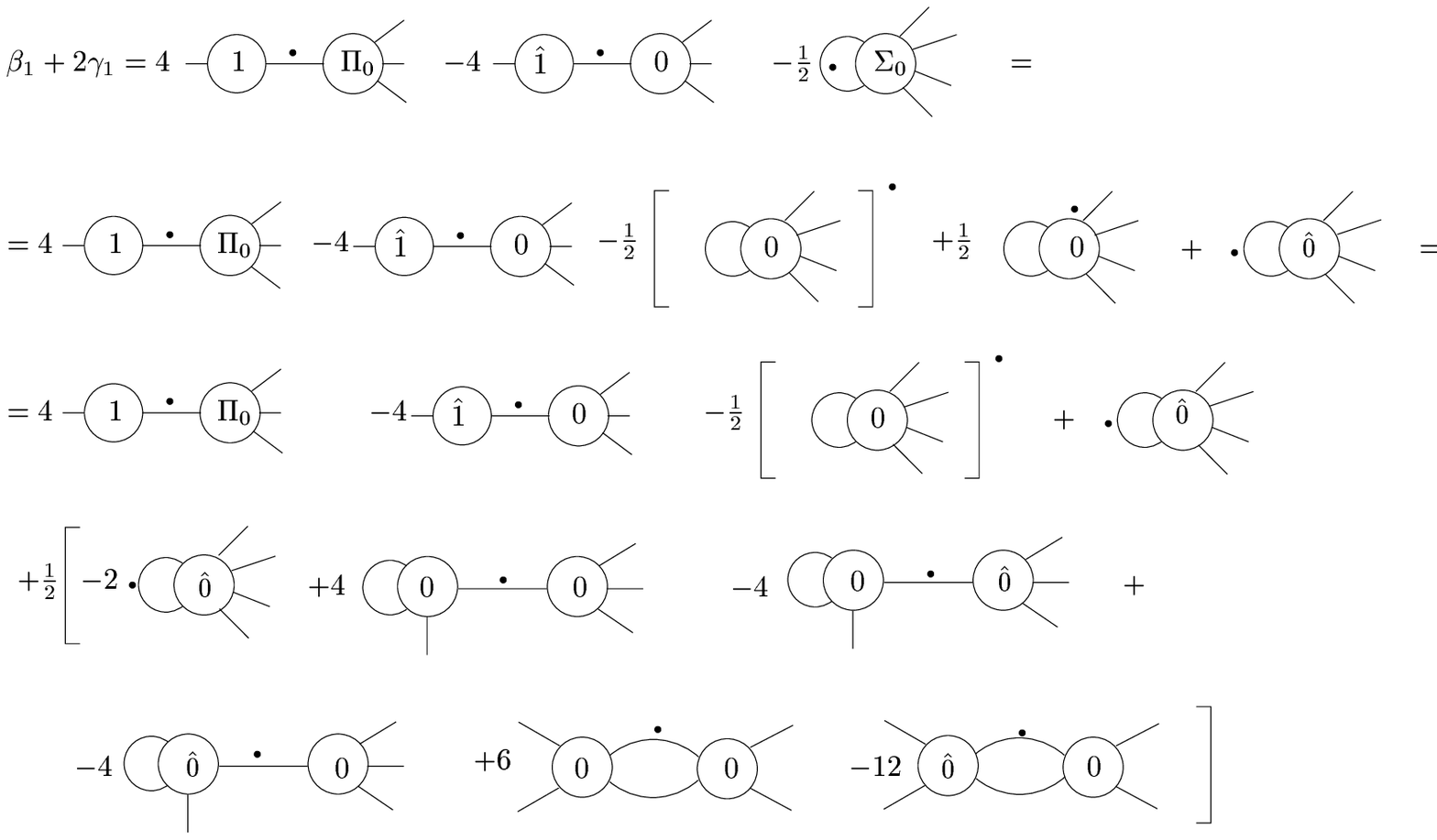}
\caption{Using the flow equation for
$S_0^{(6)}$. All the external lines carry zero momentum.\label{fig:6pteq}}
\end{figure*}
\end{widetext}
Having a closer look at fig.~\ref{fig:6pteq}, we see that the
terms containing six-point $\hS$ vertices cancel out (last diagram
on the third row and first on the fourth). Moreover, the second
and third diagrams on the fourth row pair up to give a $\Pi$
four-point vertex, which in turn cancels against 
the first diagram on the third row. This is clear once this latter is
expressed in terms of tree-level vertices. 
In more detail, the
one-loop flow equation for the two-point coupling at null momentum
is just given by fig.~\ref{fig:intparts} except with $p=0$ and
without the ${\cal O}(p^2)$ restriction. With the same
manipulations we get fig.~\ref{fig:4pteq}, again without the
order $p^2$ restriction and with $p=0$ (causing the last term in
fig.~\ref{fig:4pteq} to vanish). Thus we are only left with the
total derivative in fig.~\ref{fig:intparts}, which can be
integrated immediately to 
\be \label{s210} 
\s{2}{1}{0} = - \frac{1}{2} \int_q \Delta_q \,\s{4}{0}{0,0,q,-q}, 
\ee 
with no
integration constant since in a massless theory there must be no
other explicit scale apart from the effective cutoff. The
total derivative term in fig.~\ref{fig:6pteq} vanishes as there
is no obstruction to exchanging the order of derivative and
integral signs.
\begin{figure}[!h]
\includegraphics{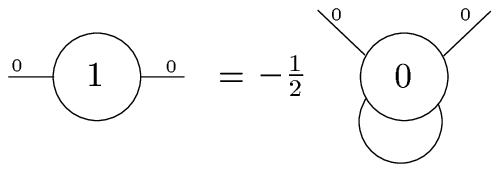}
\caption{Expressing $\s{2}{1}{0}$ in terms of
  tree-level vertices [\cf \eq{s210}].\label{fig:s210}}
\end{figure}

We still have to deal with the last three diagrams in fig.~\ref{fig:6pteq}. 
Of these, we can only ``attack'' the one with no
dependence upon $\hS$, or else we would need to deal with the
(generic) derivatives of the seed action vertices. So, integrating
by parts and using the classical flow equations, the diagrams shown in
fig.~\ref{fig:remainders} are arrived at. 
\begin{widetext}
\phantom{ABCDE} 
\begin{figure*}[!h]
\includegraphics{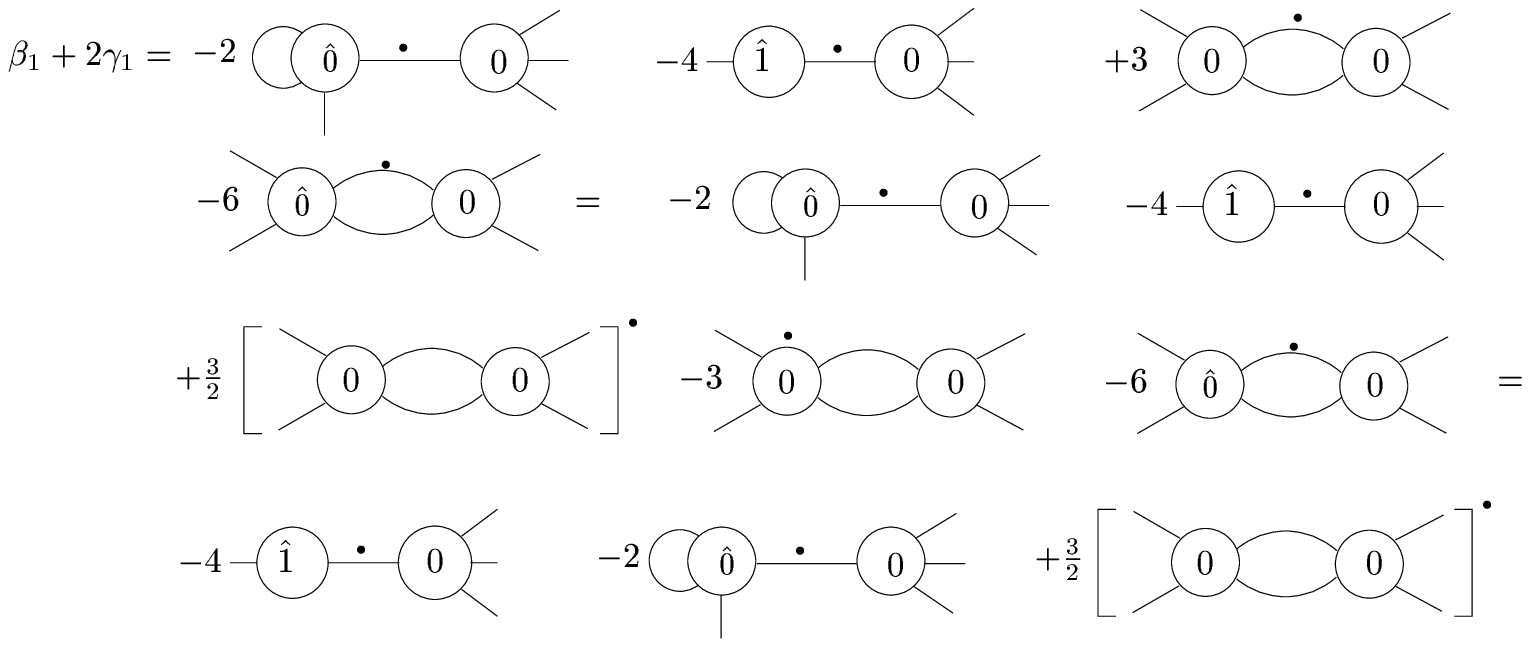}
\caption{Further
processing \eq{beta1}, from fig.~\ref{fig:6pteq}. All external
legs carry zero momentum. To integrate by parts the second
diagram, we use the fact that $\Delta_q \dd_q = \frac{1}{2}
(\Delta_q^2)^{\scriptscriptstyle \bullet
}$.\label{fig:remainders}}
\end{figure*}
\begin{figure*}[!h]
\includegraphics{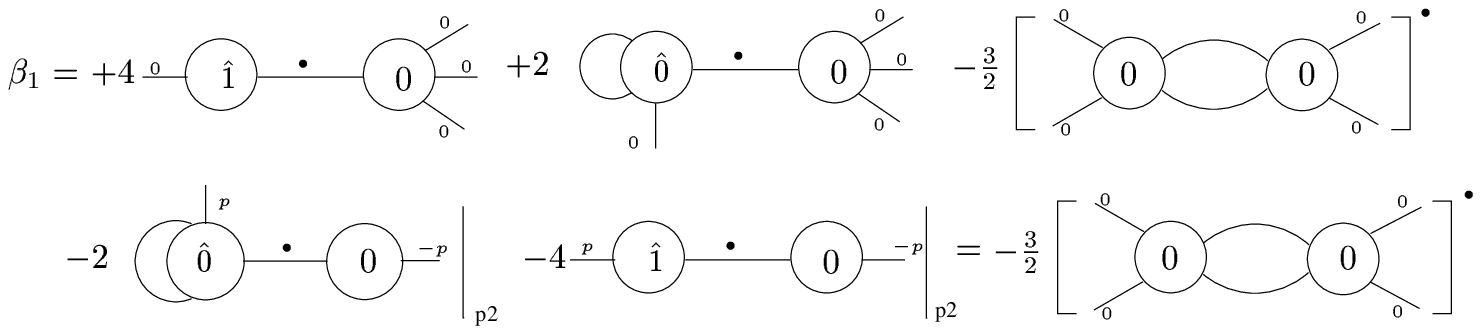}
\caption{Diagrammatic representation of
  $\beta_1$.\label{fig:last}}
\end{figure*}
\end{widetext}
Collecting our results and solving for $\beta_1$ we get just the five
diagrams in fig.~\ref{fig:last}, of which only the third remains after
the renormalization conditions are taken into account. 
In fact, as
$\s{4}{0}{\vec{0}} = \left. \s{2}{0}{p} \right|_{p^2} = 1$, the algebraic
expressions corresponding to the other four
diagrams cancel out in pairs.

In formulae,
\bea  \label{beta1last}
\beta_1&=&\frac{3}{2} \int_q \frac{1}{q^4}\, \ldl\, \Big\{ c_q
\s{4}{0}{0,0,q,-q} \Big\}^2  \nonumber\\[3pt]
&=& -\frac{3}{2}{\Omega_4\over(2\pi)^4}
\int_0^{\infty}\!\!\! dq \, \de_q \left\{ c_q
\, \s{4}{0}{0,0,q,-q} \right\}^2 \nonumber\\[3pt]
&=&\frac{3}{16\pi^2},
\eea
which is the standard one-loop result \footnote{The term in braces depends
only on $q^2/\Lam^2$. $\Omega_4$
is the four dimensional solid angle. The last line follows from the
convergence of the integral and
normalization conditions, $c(0)=1$ and $\s{4}{0}{\vec{0}} = 1$.}.

\section{Two-loop beta function}

This section is devoted to computing the two-loop beta function with a
generalised seed action. The same procedure as in the previous section will
be followed, namely:\\[0.2cm]
- introduce integrated kernels in the diagrams containing effective action
  vertices only and integrate by parts
  so as to end up with a total $\Lam$-derivative contribution plus
terms where the
$\Lam$-derivative acts on the effective action vertices 
(\cf fig.~\ref{fig:intparts});\\[0.2cm] 
- make use of the flow equation to process these latter
  further (\cf fig.~\ref{fig:4pteq});\\[0.2cm]
- use the relation between the tree-level two-point vertex
and the integrated kernel, \eq{intwin}, {\it i.e.\ \emph{fig.~\ref{fig:intwin}}}, 
to simplify the structure of the diagrams. At
this point, many cancellations should become evident;\\[0.2cm]
- repeat the above procedure when dealing with the
diagrams generated by the use of the flow equations, until there 
is no dependence at all upon the (generic) seed action.\\[0.2cm]
The one-loop relations, eqs.~(\ref{beta1}), (\ref{gamma1}), as well as
\eq{s210} will also be used to recast some terms in a more convenient form.

\begin{figure}[!h]
\includegraphics{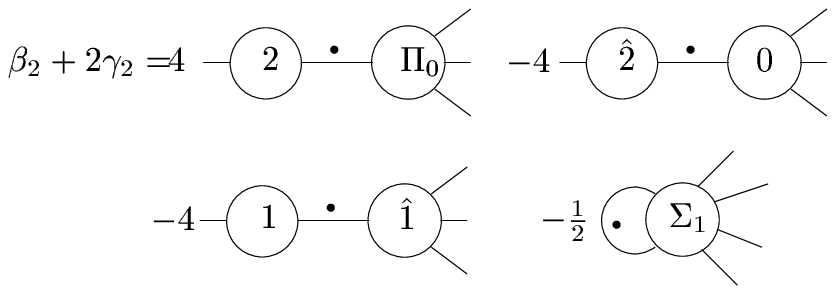}
\caption{Graphical representation of \eq{beta2}.
All the external legs have null momentum.\label{fig:beta2}}
\end{figure}

\begin{figure}[!h]
\includegraphics{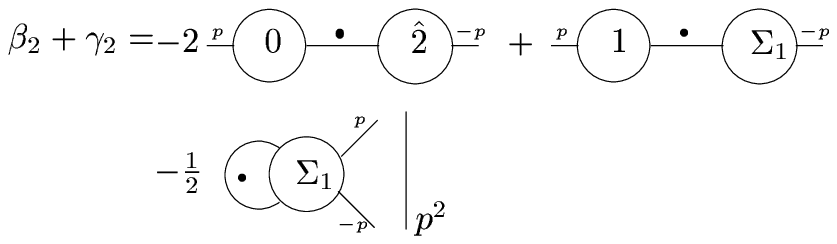}
\caption{Graphical
representation of \eq{gamma2}.\label{fig:gamma2}}
\end{figure}

Moving on to the actual calculation, we start off by writing the
RG equations for $\s{4}{2}{\vec{0}}$ and $\left.\s{2}{2}{p}
\right|_{p^2}$, which reduce to purely algebraic relations when the
renormalization conditions are imposed. They take the form:
\bea
\beta_2+\!\!&\!2\!&\!\!\gamma_2 = 4 \s{2}{2}{0} \, \dd_0 \, \Pi_0^{(4)}
(\vec{0}) - 4 \hs{2}{2}{0} \, \dd_0 \, \s{4}{0}{\vec{0}}\nonumber\\ 
&\!-\!& 4 \s{2}{1}{0} \, \dd_0 \, \hs{4}{1}{\vec{0}}
- \frac{1}{2}\!\int_q \dd_q \sig{6}{1}{\vec{0},q,-q},\label{beta2} 
\eea
\bea
\beta_2+\gamma_2 &\ug & \left. -2 \s{2}{0}{p} \, \dd_p \, 
\hs{2}{2}{p}\right|_{p^2}
 + \left. \s{2}{1}{p} \, \dd_p \, \sig{2}{1}{p} \right|_{p^2} \nonumber\\
&-& \frac{1}{2} \!\int_q  \dd_q \,
\left. \sig{4}{1}{p,-p,q,-q}\right|_{p^2}, \label{gamma2} 
\eea

and are represented diagrammatically as in figs.~\ref{fig:beta2},
\ref{fig:gamma2}. 
\ \\

It is easy to see that the algebraic expressions for the diagrams
containing the seed action vertices at the highest possible loop order, \ie
the second in fig.~\ref{fig:beta2} and the first in fig.~\ref{fig:gamma2}, 
cancel out when solving for $\beta_2$ once the renormalization 
conditions, \eq{rencon}, have been 
taken into account [\cf the comment below fig.~\ref{fig:remainders}]. 
Therefore, in what follows, those two diagrams will not appear in the
graphical representations of eqs.~(\ref{beta2}), (\ref{gamma2}).

As in the one-loop case, we will start with the easier part, \ie \eq{gamma2}.
While the first two terms in fig.~\ref{fig:gamma2} need just 
expanding at order
$p^2$, the third has to be fully processed. 

Introducing the integrated
kernel and making use of the flow equation for $\s{4}{1}{p,-p,q,-q}$ we get
the contributions in fig.~\ref{fig:2loop4pteq}.
Of these, we can only further simplify v and vi, by unfolding
the four- and six-point tree-level couplings together with the
two-point one-loop vertex. The way forward is essentially the same
as before, but with one {\it caveat}. Integrating by parts, that
is trading  $\dd_q \sim 1/\Lam^2$ for $\Delta_q \sim 1/q^2$, might
cause the appearance of infrared divergences, absent in the first
place as the ERG equation is well defined in the infrared. Of
course, these divergences cancel out when all the diagrams on the
r.h.s. are taken into account, but individual contributions might
be infrared divergent. As an example, let us go back 
to fig.~\ref{fig:2loop4pteq} 
and focus on v \footnote{It should be 
evaluated at order $p^2$, but divergences cancel at any order.}.

\begin{widetext}
\phantom{ABCDE}
\begin{figure*}[!h]
\includegraphics{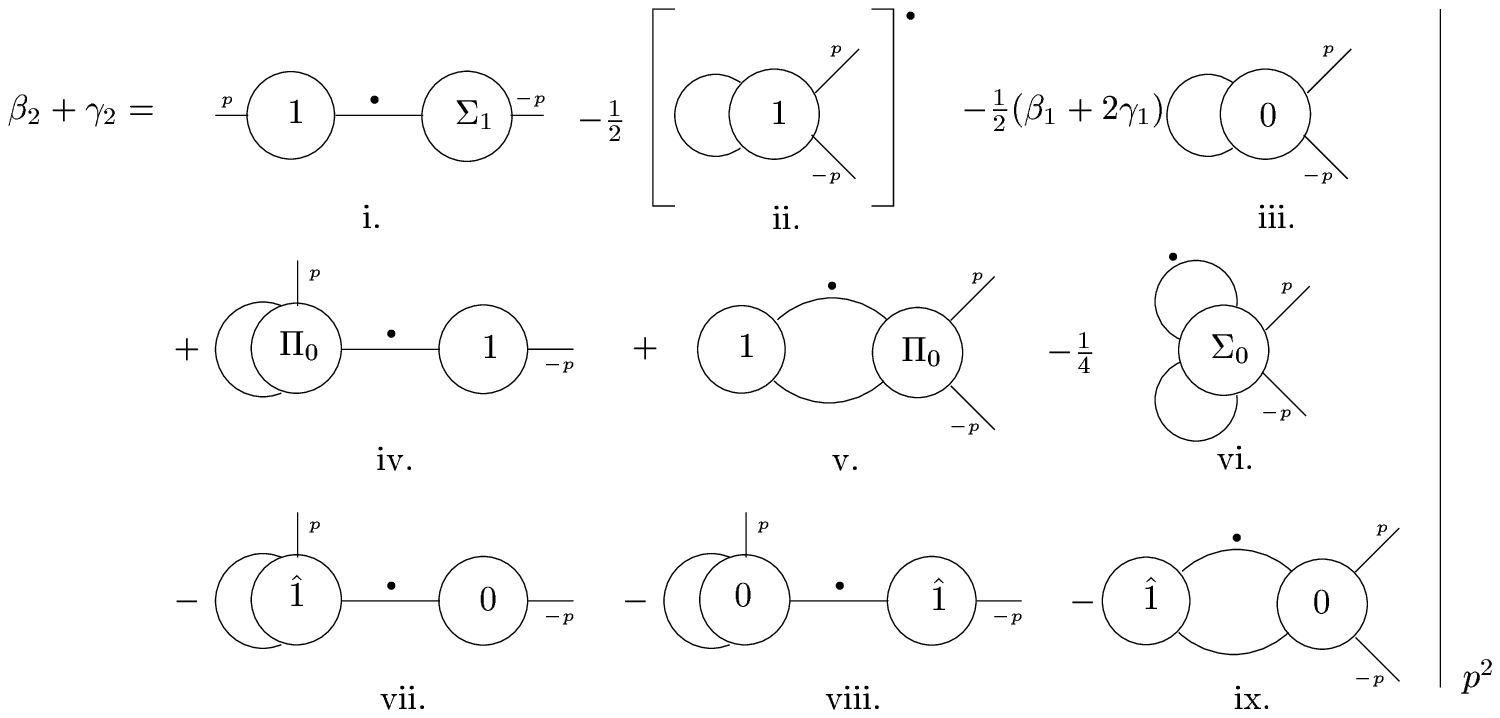}
\caption{Unfolding the one-loop four-point
  effective coupling.\label{fig:2loop4pteq}}
\end{figure*}
\end{widetext}

Upon integration by parts we obtain the contributions shown in 
fig.~\ref{fig:rearrange}.

Of these, the last two are clearly divergent. However their divergences
must cancel out by construction, as the diagram on the l.h.s.\ of fig.~\ref{fig:rearrange} 
is infrared finite.
In the case of the previous example, we could have
written $\s{2}{1}{q} = [\s{2}{1}{q}]_R + \s{2}{1}{0}$, which defines the
reduced vertex, $[\s{2}{1}{q}]_{R}$. The reduced vertex is 
at least ${\cal O}(q^4)$, and this is what we could have processed further.
All the
diagrams obtained by integration by parts would have been finite,
containing either a reduced vertex or a $\dd$.
For simplicity's sake, we will continue with the original method, bearing
in mind that all infrared divergences indeed cancel out, which we will
verify at the end of the calculation.

\begin{widetext}
\phantom{ABCDE}
\begin{figure*}[!h]
\includegraphics{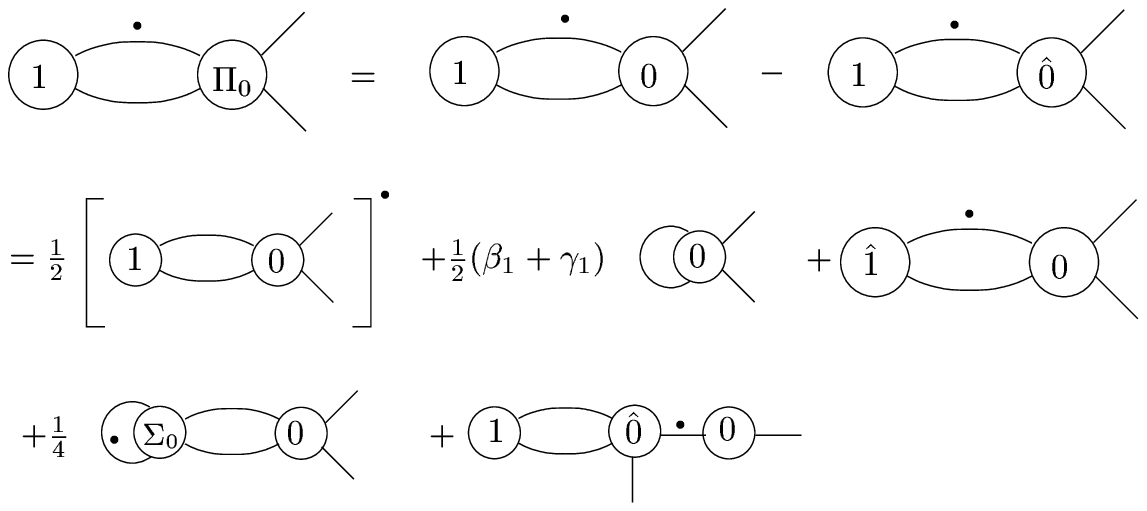}
\caption{Appearance of infrared
  divergences. External legs respectively carry $p$ and
  $-p$.\label{fig:rearrange}} 
\end{figure*}
\end{widetext}

Returning to fig.~\ref{fig:2loop4pteq} 
and iterating our procedure, we obtain the diagrams 
in fig.~\ref{fig:b2g2}.

\begin{widetext}
\phantom{ABCDE}
\begin{figure*}[!h]
\includegraphics{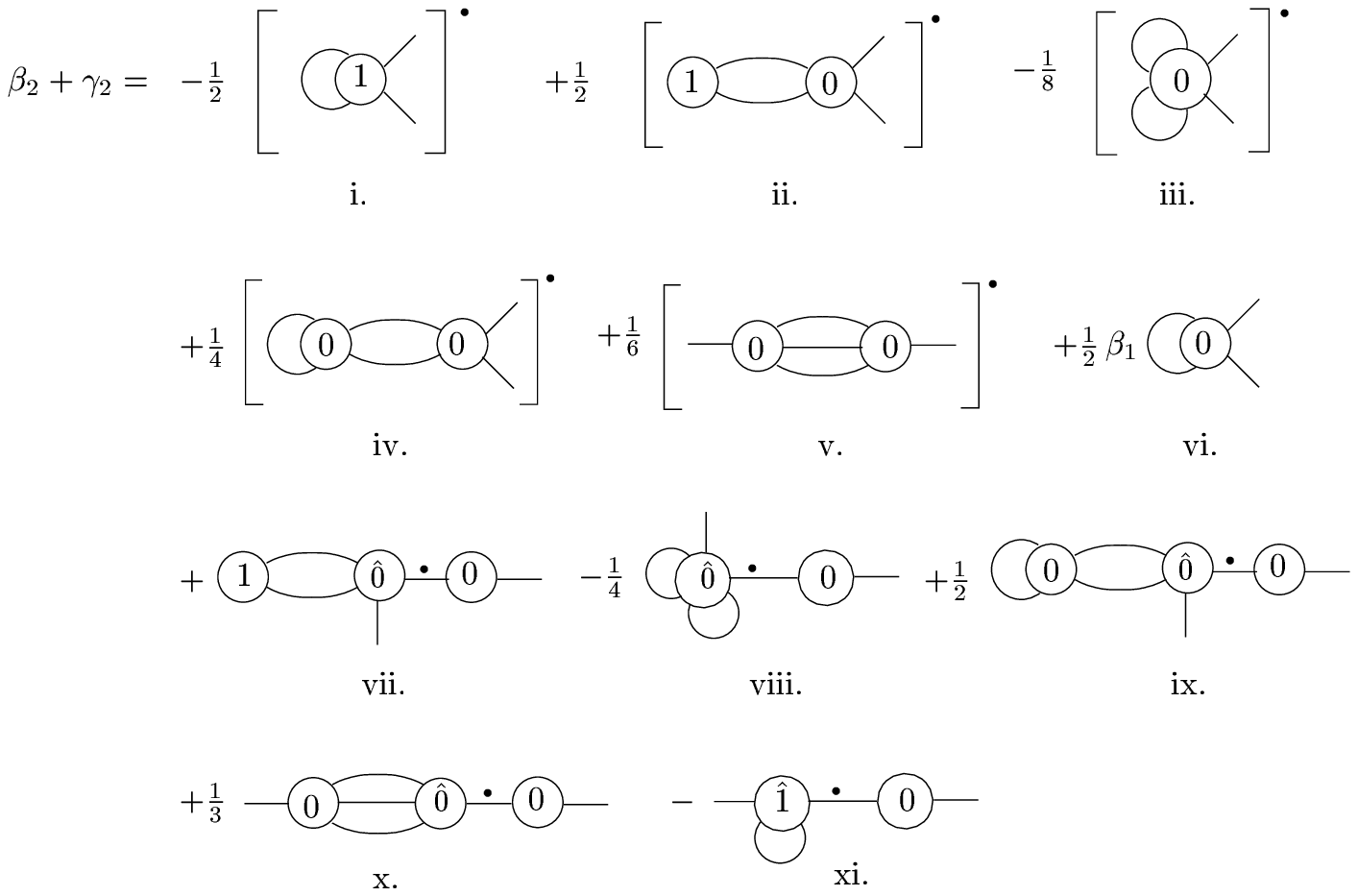}
\caption{\ceq{gamma2} in its final
  form. External legs carry momenta $p$ and $-p$ respectively. The r.h.s.\
  must be evaluated at ${\cal O}(p^2)$.\label{fig:b2g2}}
\end{figure*}
\end{widetext}

Of those, i and iii vanish identically as they are (ultraviolet-
and) infrared-finite dimensionless integrals, so do ii and iv
after their infrared-divergent parts have been cancelled against
each other by means of fig.~\ref{fig:s210} [\cf \eq{s210}].
(N.B. Neither ii nor iv is infrared divergent after the
$\Lam$-derivative is taken, however they each yield a finite
non-zero answer. The strategy is that by combining the two
diagrams we get one that is finite before differentiation, and
thus zero afterwards.) We are then left with v, the wavefunction
renormalization contribution to the beta function; vi, which will
cancel a logarithmic divergence under a total $\ldl$, and with the
last five diagrams, vii - xi, which all contain $\s{2}{0}{p}$.
Their corresponding algebraic expressions will cancel against
similar contributions from \eq{beta2} 
by the renormalization conditions, in exactly the same fashion as
described in fig.~\ref{fig:last}. Diagram vi is arrived at by combining
the expansion of i, iii, iv, viii in fig.~\ref{fig:2loop4pteq} at order
$p^2$ together with that of other two
diagrams coming from vi in the same figure and using fig.~\ref{fig:b1g1mod}.

Let us now deal with \eq{beta2}. In
complete analogy with the one-loop calculation, we first solve the flow
equation for the two-point coupling at null momentum. The two-loop analogue
of \eq{s210} is shown in fig.~\ref{fig:s220}.
(As before, the first two diagrams are actually infrared divergent,
but the divergent parts cancel out if \eq{s210} is taken into account.)

Next, we process the second diagram in fig.~\ref{fig:beta2}. After
several iterations, we arrive at the
final form for \eq{beta2}, shown in fig.~\ref{fig:b22g2}.

\begin{widetext}
\phantom{ABCDE}
\begin{figure*}[!h]
\includegraphics{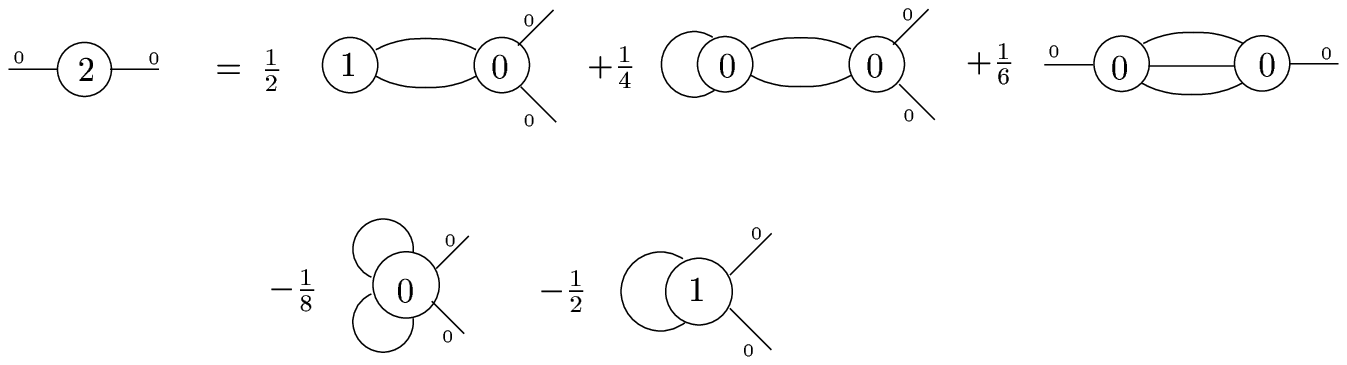}
\caption{Diagrammatical solution of the flow
  equation for $\s{2}{2}{0}$.\label{fig:s220}}
\end{figure*}

\begin{figure*}[!h]
\includegraphics{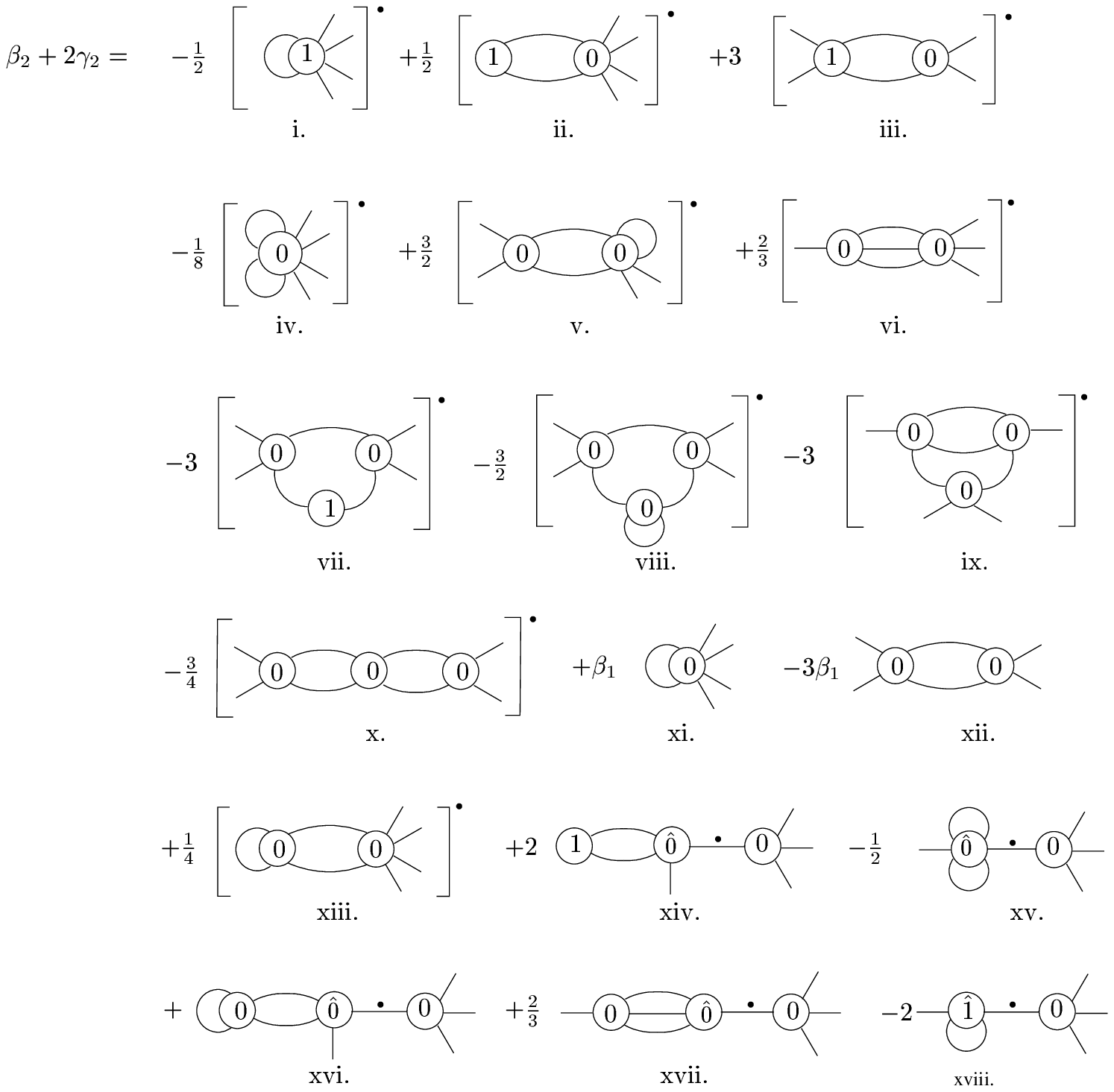}
\caption{\ceq{beta2} in its final
  form. External legs carry zero momenta.\label{fig:b22g2}}
\end{figure*}
\end{widetext}
Again, there are many cancellations. In detail, i, iii, iv and vi
vanish
identically as they are convergent and dimensionless \footnote{Remember
$\s{4}{1}{\vec{0}}$ is zero by the renormalization condition, so
there is no infrared divergence in iii.}, and so do ii and xiii
after their infrared-divergent parts have been cancelled against each
other. Moreover, the infrared-divergent part of v results in the finite
contribution $- \beta_1 \int_k
\s{6}{0}{\vec{0},k,-k}\, \Delta_k$, which cancels xi exactly.
In a somewhat similar fashion, we see that the divergent part \footnote{In
this case, divergent even after performing $\ldl$.} of
viii is equal and
opposite to that of vii, up to a term that will cancel vi in
fig.~\ref{fig:b2g2}, \ie
\be
(\mbox{viii})_{\mbox{div}} = - (\mbox{vii})_{\mbox{div}} +
\beta_1 \int_k \left. \s{4}{0}{q,-q,k,-k} \, \Delta_k \right|_{q^2}.
\ee
(In more detail, by the renormalization conditions $\s{2}{1}{q}$ has no
${\cal O}(q^2)$ part, its zeroth order part being that of fig.~\ref{fig:s210}. 
On the contrary, 
the ${\cal O}(q^2)$ part in viii remains uncancelled.) Diagrams xi and xii
again result from pairing up several topologically different diagrams 
and using fig.~\ref{fig:b1g1mod}.

As for the last five diagrams, xiv - xviii, they respectively
cancel the last five terms in fig.~\ref{fig:b2g2} when the two equations are
solved for $\beta_2$. Hence, the only diagrams to be evaluated are those in
fig.~\ref{fig:b2}.
\begin{widetext}
\phantom{ABCDE}
\begin{figure*}[!h]
\includegraphics{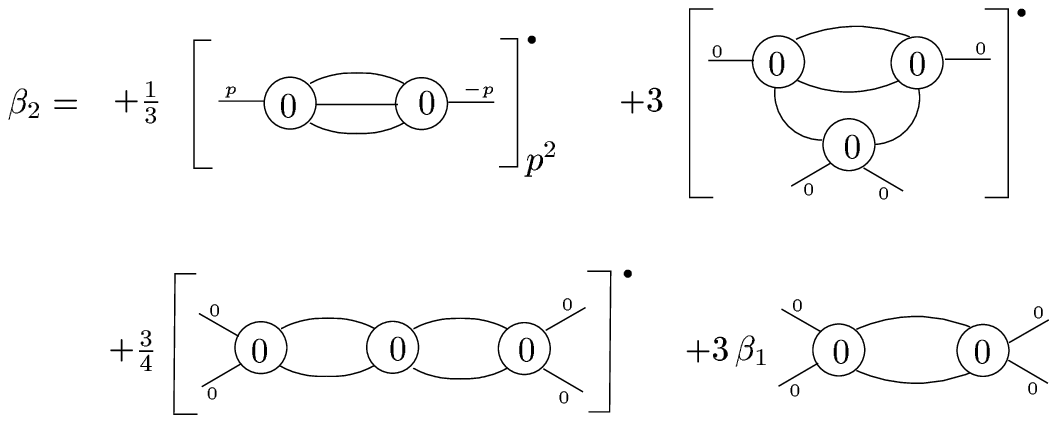}
\caption{Contributions to $\beta_2$.\label{fig:b2}}
\end{figure*}
\end{widetext}

The first three are already expressed as total derivatives, whereas the
fourth can be turned into one by means of 
\eq{beta1last} (\cf fig.~\ref{fig:2looplast}).
\begin{widetext}
\phantom{ABCDE}
\begin{figure*}[!h]
\includegraphics{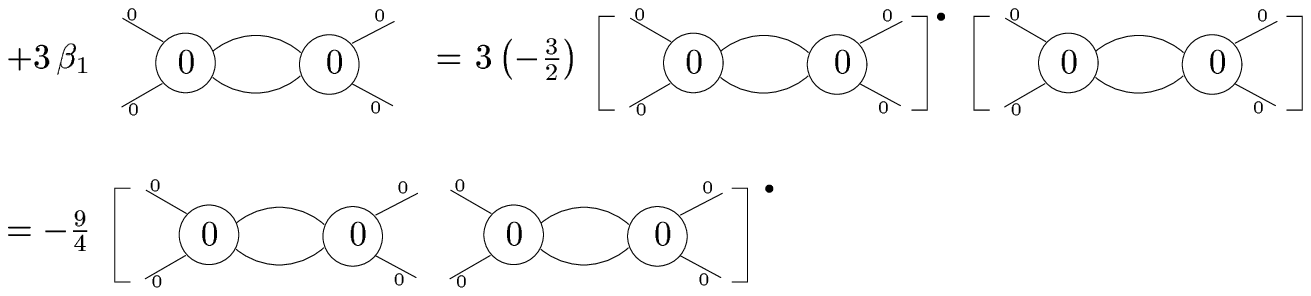}
\caption{Turning last term in fig.~\ref{fig:b2} 
into a total derivative.\label{fig:2looplast}}
\end{figure*}
\end{widetext}
Although trading a numerical coefficient, $\beta_1$,
for a diagram might be regarded as an added complication, it actually
greatly simplifies the rest of the calculation, as we now just need to
extract the infrared part of each diagram.
In order to do this, the reduced four-point vertices are introduced,
which by construction (and Lorentz invariance) vanish at least as the following
powers of momentum:
\be
\bear
{\ds S_{\! R} \, (k) \doteq \s{4}{0}{0,0,k,-k} -
\s{4}{0}{\vec{0}} = {\cal O}(k^2), }\nonumber\\[0.3cm]
{\ds S_{\! R} \, (k,q) \doteq
\s{4}{0}{k,-k,q,-q} - \s{4}{0}{k,-k,0,0} }\nonumber\\[0.3cm]
{\ds
- \s{4}{0}{0,0,q,-q} +
\s{4}{0}{\vec{0}} = {\cal O}(q\cdot k).   }
\eear
\ee

Eliminating all four-point vertices in favour of the reduced ones,
we can easily single out the infrared-divergent parts, discarding
finite contributions which would vanish after the derivative with
respect to $\Lam$ is taken.

The third term in fig.~\ref{fig:b2}, for example, can be written as

\be \label{terzoequarto}
\bear
{\ds \frac{3}{4} \int_{k,q} \! \s{4}{0}{0,0,q,-q} \, \Delta_q^2 \,
  \s{4}{0}{q,-q,k,-k} }\nonumber\\[0.3cm]
{\ds \times  \Delta_k^2 \, \s{4}{0}{0,0,k,-k} }\nonumber\\[0.3cm]
{\ds = \frac{3}{4} \int_{k,q} \! \Big(S_{\! R} \,(q) +1\Big) \, \Delta_q^2
  \, \Big(S_{\! R} \,(k,q) + S_{\! R} \,(k) }\nonumber\\[0.3cm]
{\ds + S_{\! R} \,(q) +1 \Big) \, \Delta_k^2
  \, \Big(S_{\! R} \,(k) +1\Big) }\nonumber\\[0.3cm]
{\ds = \frac{3}{4} \int_{k,q} \Big[ 2 \Big( S_{\! R} \,(q) +1\Big)^2 -1\Big]
  \, \Delta_q^2 \, \Delta_k^2 }\nonumber\\[0.3cm]
{\ds =\frac{3}{4} \int_{k,q} \Big[ 2 S^2(q) -1\Big]
  \, \Delta_q^2 \, \Delta_k^2,}
\eear
\ee

where $S(q)$ stands for (the un-reduced) $\s{4}{0}{0,0,q,-q}$.

The diagram in fig.~\ref{fig:2looplast} has exactly the same structure, but
with $- \frac{9}{4}$ in front. The second term in fig.~\ref{fig:b2} can 
be simplified to

\be \label{secondo}
3 \int_{k,q} \! \Big[ S(q)^2 \, \Delta_k^2 \,\Delta_q \, \Delta_{k+q}
\Big]^{\scriptscriptstyle \bullet}.
\ee

Putting the three together yields

\be \label{together}
\bear
{\ds 3 \int_{k,q} \! \Big[ S(q)^2 \, \Delta_k^2 \,\Delta_q \Big( \Delta_{k+q} -
  \Delta_q \Big) \Big]^{\scriptscriptstyle \bullet} + \frac{3}{2}
\int_{k,q} \! \Big[ \Delta_q^2 \,\Delta_k^2 \Big]^{\scriptscriptstyle
  \bullet} }\\[0.35cm]
{\ds = 3 \int_{k,q} \! \Big[ \Delta_k^2 \,\Delta_q \Big( \Delta_{k+q} -
  \Delta_q \Big) \Big]^{\scriptscriptstyle \bullet} + \frac{3}{2}
\int_{k,q} \! \Big[ \Delta_q^2 \,\Delta_k^2 \Big]^{\scriptscriptstyle
  \bullet} }\\[0.35cm]
{\ds = 3 \int_{k,q} \! \Big[ \Delta_k^2 \,\Delta_q \Big( \Delta_{k+q} -
  \frac{1}{2} \Delta_q \Big) \Big]^{\scriptscriptstyle \bullet}, }\\
\eear 
\ee 
where we again used $S(q)=S_R(q)+1$ to discard the
contributions that are infrared finite before differentiation with
respect to $\Lam$. As regards the first diagram in fig.~\ref{fig:b2}, 
it amounts to

\be\label{primo}
\bear
{\ds \frac{1}{3} \int_{k,q} \! \Big[ \left(\s{4}{0}{-p,k,q+p,-k-q}\right)^2
\Delta_k \,\Delta_{q+k} \, \Delta_{q+p} \Big]^{\scriptscriptstyle
    \bullet}_{p^2} }\\[0.35cm]
{\ds = \frac{1}{3} \int_{k,q} \! \Big[ \left(\s{4}{0}{0,k,q,-k-q}\right)^2
\Delta_k \,\Delta_{q+k} \, \Delta_{q+p} \Big|_{p^2}
\Big]^{\scriptscriptstyle \bullet} }\\[0.35cm]
{\ds = \frac{1}{3} \int_{k,q} \! \Big[
\Delta_k \,\Delta_{q+k} \, \Delta_{q+p} \Big|_{p^2}
\Big]^{\scriptscriptstyle \bullet}, }
\eear
\ee
as the only non-vanishing contribution comes from taking the order $p^2$
from one of the propagators.

The integrals (\ref{together}),(\ref{primo}),
representing
the usual contributions to the two-loop beta function,
can be easily computed for a generic cutoff
 function
if Bonini's lead~\cite{Bo} is followed. It consists in ``eliminating''
one of the
cutoff functions by writing, say, $c_q = (c_q -1) +1$ and neglecting the
first term as it is already of order $q^2$.
This, of course, can only be done if
the integrand remains UV regulated afterwards and if
the ${\cal O}(q^2)$ part indeed makes the integral infrared convergent.

Taking \eq{primo} as an example, we can trade $c_k$ for $1$, \eg
\be
\bear
{\ds \frac{1}{3} \int_{k,q} \! \Big[ \Delta_k \,\Delta_{q+k} \,
\Delta_{q+p} \Big|_{p^2} \Big]^{\scriptscriptstyle \bullet} =
\frac{1}{3} \int_{k,q} \! \Big[ \frac{1}{k^2} \,\Delta_{q+k} \,
\Delta_{q+p} \Big|_{p^2} \Big]^{\scriptscriptstyle \bullet} 
}\nonumber\\[0.35cm]
{\ds =
\frac{2}{3} \int_{k,q} \! \frac{1}{(k-q)^2} \, \Delta_k \left(
\frac{2}{\Lam^2} c'_{q+p} \right)_{p^2}, }
\eear
\ee 
where the last
equality follows from taking the derivative with respect to $\Lam$
and shifting $k \rightarrow k-q$.

Averaging over the angles and taking the order ${p^2}$, we arrive at the
final result (the calculation is detailed in the appendix),
\be
\frac{2}{3} \int_{k,q} \! \frac{1}{(k-q)^2} \, \Delta_k
\left( \frac{2}{\Lam^2} c'_{q+p} \right)_{p^2}
= \frac{1}{3} \left(\frac{1}{16 \pi^2} \right)^2.
\ee

The integral in \eq{together} can be dealt with in pretty much the same
way (for the details see the appendix):
\be
\bear
{\ds 3 \int_{k,q} \! \Big[ \Delta_k^2 \,\Delta_q \Big( \Delta_{k+q} -
  \frac{1}{2} \Delta_q \Big) \Big]^{\scriptscriptstyle \bullet} =
3 \int_{k,q} \frac{(c_k^2 \, c_q^2 )^{\scriptscriptstyle \bullet}}{k^4 \,
  q^2} }\nonumber\\[0.35cm]
{\ds \times
\left( \frac{1}{(k+q)^2}-\frac{1}{2 q^2} \right) = -6
 \left(\frac{1}{16 \pi^2} \right)^2,  }
\eear
\ee
and summing up the two contributions the standard result follows:
\be
\beta_2 = -\frac{17}{3} \left(\frac{1}{16 \pi^2} \right)^2.
\ee


\acknowledgments We acknowledge financial support from PPARC
Rolling grant PPA/G/O/2000/00464 (TRM, SA),  PPARC SPG
PPA/G/S/1998/00527 (AG) and the Southampton University Development
Trust (OJR).

\appendix

\section{The evaluation of two-loop integrals}

This appendix is devoted to the detailed calculation of
(\ref{together}), (\ref{primo}).

Let us consider first the wavefunction renormalization
contribution, \eq{primo}. Rewriting $c_k$ as $(c_k-1)+1$, we see
that $(c_k-1)$ already makes the integral convergent in the
infrared, thus giving a vanishing contribution when the derivative
with respect to $\Lam$ is taken. Therefore we only retain the $1$:
\be
\frac{1}{3} \int_{k,q} \! \Big[
\Delta_k \,\Delta_{q+k} \, \Delta_{q+p} \Big|_{p^2}
\Big]^{\scriptscriptstyle \bullet}
= \frac{1}{3} \int_{k,q} \! \Big[
\frac{1}{k^2} \,\Delta_{q+k} \, \Delta_{q+p} \Big|_{p^2}
\Big]^{\scriptscriptstyle \bullet}.
\ee
Taking the $\Lam$-derivative and bearing in mind the permutation symmetry
in the three momenta:
\be
\frac{2}{3} \int_{k,q} \!
\frac{1}{(k-q)^2} \, \Delta_k \left( \frac{2}{\Lam^2} c'_{q+p} \right)_{p^2},
\ee
where we have also shifted $k$ to $k-q$.
Defining $\theta_k$ to be the angle between the Euclidean
$4$-vectors \footnote{The same symbol is used for the
  $4$-vector and its modulus. It should hopefully be clear from the context
what we mean by it.} $k$ and
$q$, \ie $k \cdot q \doteq k \, q \cos \theta_k$, we can perform the angular
integration in $k$ to get
\be
\bear
{\ds \frac{4}{3 \Lam^2} \int_{q} c'_{q+p} \Big|_{p^2}
\int_0^\infty  \!\! d k
\, k^3 \Delta_k \int \!\! \frac{d \Omega_k}{(2 \pi)^4}\frac{1}{(k-q)^2}
}\nonumber\\ 
{\ds 
= \frac{4 \osl_4}{3 \Lam^2} \int_{q} c'_{q+p} \Big|_{p^2}
\int_0^\infty  \!\! d k
\frac{k \, c_k}{\max\{k^2,q^2\}},   }
\eear
\ee
where $\osl_4$ is the four-dimensional solid angle divided by $(2 \pi)^4$.
Expanding $c'_{q+p}$ to take the order $p^2$ and defining $\theta_q$ as the
angle between $q$ and $p$, we can integrate over the solid angle, $d \Omega_q$,
\be
\frac{4 \osl^2_4}{3 \Lam^4} \int_0^\infty \!\! d q \, q^3
\left(c''_q+\frac{q^2}{2\Lam^2}c'''_q\right) \left\{
\int_0^q  \!\! d k
\frac{k \, c_k}{q^2} + \int_q^\infty \!\! d k
\frac{c_k}{k} \right\}.
\ee
We now introduce two dimensionless variables, $x = \frac{k^2}{\Lam^2}, y =
\frac{q^2}{\Lam^2}$ and recast the above as
\be \label{domains}
\frac{\osl^2_4}{3} \int_0^\infty \!\!\! d y
\left(c''_y+\frac{y}{2}c'''_y\right) \left\{
\int_0^y  \!\! d x \, c_x + \int_y^\infty\! \!\! d x \,
y \, \frac{c_x}{x} \right\},
\ee
with $c^{(n)}_{x}$ being $c^{(n)}(x)$ and similarly for $y$.

Exchanging the order of integration, so that the integral over $y$ is
performed first, and temporarily discarding the numerical factor in 
front of the integral
\be
\bear
{\ds 
\int_0^\infty\!\!\!\!\! d x \, c_x \left\{ \int_x^\infty
\!\!\!\!\! d y \, \left(c''_y+\frac{y}{2}c'''_y\right) + \int_0^x
\!\!\! d y \left( \frac{y}{x} \, c''_y+\frac{y^2}{2x}c'''_y\right) \right\}
=}\nonumber\\[0.4cm]
{\ds 
\left\{ -\frac{1}{2} x \, c_x \, c'_x
  \Big|_0^\infty + \frac{1}{2} \int_0^\infty \!\! dx\,  x \, (c'_x)^2
+ \int_0^\infty \!\!\! dx\,  \frac{c_x}{x} \frac{x^2}{2} c''_x \right\}=
}\nonumber\\[0.4cm]
{\ds  
\left( -\frac{1}{4} c^2_x \right)_0^\infty = \frac{1}{4}.
}
\eear
\ee

Hence, \eq{primo} amounts to
\be
\frac{1}{4} \frac{\osl^2_4}{3} = \frac{1}{3} \left(\frac{1}{16 \pi^2}
\right)^2.
\ee

As far as \eq{together} is concerned, we can follow the same strategy and
write $c_{k+q} = (c_{k+q} - c_q) + c_q$, retaining only $c_q$:
\be
\bear
{\ds 
3 \int_{k,q} \! \Big[ \Delta_k^2 \,\Delta_q \Big( \Delta_{k+q} -
  \frac{1}{2} \Delta_q \Big) \Big]^{\scriptscriptstyle \bullet} 
}\nonumber\\[0.4cm]
{\ds =
3 \int_{k,q} \frac{(c_k^2 \, c_q^2 )^{\scriptscriptstyle \bullet}}{k^4 \,
  q^2} \left( \frac{1}{(k+q)^2}-\frac{1}{2 q^2} \right).   }
\eear
\ee

Averaging over the angles and rewriting in terms of $x,y$ we get
\bea \label{ant}
\frac{3}{2} \osl^2_4\!\! &&\Bigg\{\intoi dx\intox dy \bigg[\frac{c^2_x}{x^2}\ y
(c^2_y)'+\frac{(c^2_x)'}{x}\ c^2_y\bigg]\nonumber\\[0.35cm]
         &&+\frac{1}{2}\intoi dx\intxi dy \bigg[\frac{c^2_x}{x}\
(c^2_y)'+(c^2_x)'\ \frac{c^2_y}{y}\bigg]\nonumber\\[0.35cm]
         &&-\frac{1}{2}\intoi dx\intox dy \bigg[\frac{c^2_x}{x}\
(c^2_y)'+(c^2_x)'\ \frac{c^2_y}{y}\bigg]\Bigg\}.
\eea

Again exchanging the order of integration and using the fact that
the integrand is invariant under the
exchange $x \leftrightarrow y$, we see the last two lines in \eq{ant} are equal and
opposite, while the first can be rewritten as
\be
\bear
{\ds 
\frac{3}{2} \osl^2_4 \left[\frac{c_x^2}{x} \intox dy\,
  c^2_y\right]_{x=0}^{\infty} = -\lim_{x\to 0}
\frac{3}{2}\osl^2_4 \frac{c_x^2}{x} \intox dy\,
  c^2_y 
}\nonumber\\[0.35cm]
{\ds 
= - \frac{3}{2}\osl^2_4 = -6 \left(\frac{1}{16 \pi^2} \right)^2. }
\eear 
\ee
\phantom{A}

\end{document}